\documentclass[11pt]{article}
\usepackage{moriond,epsfig}

\def\be{\begin{equation}}
\def\ee{\end{equation}}
\def\bea{\begin{eqnarray}}
\def\eea{\end{eqnarray}}

\begin{document}
\vspace*{4cm}
\title{SOLVING DEGENERACIES AT A $\beta$-BEAM BY COMBINING IONS}

\author{E. FERNANDEZ-MARTINEZ}

\address{I.F.T. and Dep. F\'{\i}sica Te\'{o}rica, Univ. Autonoma Madrid., E-28049, Madrid, Spain}

\maketitle\abstracts{
We study how the eightfold-degeneracy in the ($\theta_{13},\delta$) plane observed in $\gamma \sim 100$ $\beta$-beams
can be reduced by periodically changing  the ions in the storage ring. This ``ions cocktail'' allows to change
the neutrino energy, at fixed $\gamma$, by choosing ions with different decay energies. We propose to combine the
standard $^6$He and $^{18}$Ne beams with $^8$Li and $^8$B ones.
These latter two ions have peaked $\nu_e \to \nu_\mu$ oscillation probabilities for $\gamma = 100$ at a baseline
$L \sim 700$ Km. At this distance and this $\gamma$ the oscillation probability of $^6$He and $^{18}$Ne neutrinos is at its second maximum.
This setup is particularly suited for large enough values of $\theta_{13}$ (within reach at T2K-I) and it allows solving
most of the eightfold-degeneracy, measuring $\theta_{13}, \delta$ and the sign of the atmospheric mass difference
for values of $\theta_{13} \geq 5^\circ$.
}

\section{The Alternating Ions Scheme}

The results of atmospheric, solar, accelerator and reactor neutrino experiments \cite{Fogli:2005cq} show
that flavour mixing occurs in the leptonic sector. Experiments have measured two mass
differences, $\Delta m^2_{sol} \approx 7.9 \times 10^{-5}$ eV$^2$ and
$|\Delta m^2_{atm}| \approx 2.4 \times 10^{-3}$ eV$^2$. But only two out of the four parameters of the three-family leptonic mixing matrix $U_{PMNS}$ are known: $\theta_{12} \approx 34^\circ$ and $\theta_{23}\approx 41.5^\circ$  \cite{Fogli:2005cq}.
The other two parameters, $\theta_{13}$ and $\delta$, are still unknown: for $\theta_{13}$ direct searches at reactors give the upper bound $\theta_{13}
\leq 11.5^\circ$, whereas for the leptonic CP-violating phase $\delta$ we have no information
whatsoever. Two additional discrete unknowns are the sign of the atmospheric mass difference and the $\theta_{23}$-octant.
The two unknown parameters $\theta_{13}$ and $\delta$ can be measured in ``appearance'' experiments
through $\nu_e \to \nu_\mu, \nu_\mu \to \nu_e$ (the ``golden channel''
\cite{Cervera:2000kp}) and $\nu_e \to \nu_\tau$ (the ``silver channel'' \cite{Donini:2002rm}) oscillations.
However, strong correlations between $\theta_{13}$ and $\delta$ and the presence of parametric degeneracies
in the ($\theta_{13},\delta$) parameter space, \cite{Burguet-Castell:2001ez}$^-$\cite{Barger:2001yr}, make
the simultaneous measurement of the two variables extremely difficult.
Most proposed solutions to these problems suggest the combination of different experiments and facilities, such as
Super-Beam's (of which T2K \cite{Itow:2001ee} is the first approved one), $\beta$-beam's \cite{Zucchelli:sa}
or the Neutrino Factory \cite{Geer:1997iz}.

Here we propose to alleviate the parametric degeneracy in the ($\theta_{13},\delta$) plane by
combining $\beta$-beam's obtained from the decay of several different ions \cite{Donini:2006dx}. The $\beta$-beam concept \cite{Zucchelli:sa} involves producing a huge number of $\beta$-unstable ions, accelerating them to some reference
energy, and letting them decay in the straight section of a storage ring, resulting in a very intense and pure
$\nu_e$ or $\bar \nu_e$ beam. ``Golden'' sub-leading transitions, $\nu_e \to \nu_\mu$ and $\bar \nu_e \to \bar \nu_\mu$,
can then be measured through muon observation in a distant detector.

\begin{table}
\begin{center}
\begin{tabular}{|c|c|c|c|c|} \hline
   Element  & $A/Z$ & $T_{1/2}$ (s) & $Q_\beta$ eff (MeV) & Decay Fraction \\
\hline
  $^{18}$Ne &   1.8 &     1.67      &        3.41         &      92.1\%    \\
            &       &               &        2.37         &       7.7\%    \\
            &       &               &        1.71         &       0.2\%    \\
  $^{8}$B   &   1.6 &     0.77      &       13.92         &       100\%    \\
\hline
 $^{6}$He   &   3.0 &     0.81      &        3.51         &       100\%    \\
 $^{8}$Li   &   2.7 &     0.83      &       12.96         &       100\%    \\
\hline
\end{tabular}
\caption{ $A/Z$, half-life and decay energy for two $\beta^+$-emitters ($^{18}$Ne and $^8$B)
and two $\beta^-$-emitters ($^6$He and $^8$Li).}
\end{center}
\label{tab:ions}
\end{table}

In Tab.~\ref{tab:ions} we remind the relevant parameters for five ions: $^{18}$Ne, $^6$He, $^8$Li and $^8$B.
Consider first $^6$He and $^{18}$Ne: as it was stressed in the literature, $^6$He
has the right half-life to be accelerated and stored such as to produce an intense $\bar \nu_e$ beam.
This is also the case for $^{18}$Ne, the best candidate as $\beta^+$-emitter. These two ions are the ones usually proposed as sources for the neutrino beams.

The two ions we propose as an alternative to $^6$He and $^{18}$Ne as $\beta^-$ and $\beta^+$-emitters are
$^8$Li and $^8$B, respectively. A novel way to produce intense beams of these ions was proposed in Ref. 12. $^8$Li and $^8$B have similar half-life, $Z$ and $A/Z$ to $^6$He and $^{18}$Ne, thus sharing the key characteristics needed
for the bunch manipulation, but both ions have a much larger end-point energy than the two reference ions.
As a consequence, for a fixed $\gamma$, a longer baseline is needed, and thus a smaller signal statistics is expected in the far detector. Therefore, the expected sensitivity to $\theta_{13}$ of such beams is smaller than that for a ``standard'' beam produced via $^6$He and $^{18}$Ne. However, when $\gamma$ is limited (for example when using the CERN SPS as it is) it is then possible to reach
higher neutrino energies using the same facility to accelerate the ions to be stored.
If we combine different ions, we can (using the same facility and the same $\gamma$ factor) produce neutrino beams of
different $L/E$ that can be used to disentangle many of the parametric degeneracies discussed before.
As it was shown in the literature (see, for example, Refs. 15-17)
degeneracies are indeed best lifted combining beams with different $L/E$.

Three classes of setups have been considered so far: $\gamma \sim 100$ (``low'' $\gamma$), with a typical baseline of
$O (100)$ Km \cite{Burguet-Castell:2003vv}$^-$\cite{Bouchez:2003fy}$^,$ \cite{Donini:2004hu}$^-$\cite{Donini:2004iv};
$\gamma \sim 300$ (``medium'' $\gamma$), with $L = O (700)$ Km \cite{Burguet-Castell:2003vv}$^,$ \cite{Burguet-Castell:2005pa}$^-$\cite{Donini:2005qg};
and $\gamma \geq 1000$ (``high'' $\gamma$), with baselines of several thousands kilometers, \cite{Burguet-Castell:2003vv}$^,$ \cite{Huber:2005jk}.

We will compare results obtained with three different setups:

{\bf Standard ``low'' $\gamma$ scenario} $L = 130$ Km (CERN to Fr\'ejus); $\gamma_{^6 He} = \gamma_{^{18} Ne} = 120$.
Both fluxes are tuned to be at the first oscillation peak.
A given ion is accumulated in the storage ring for ten years.

{\bf Alternating ions ``low'' $\gamma$ scenario} $L = 650$ Km (CERN to Canfranc);
$\gamma_{^8 Li} = \gamma_{^8 B} = 100$; $\gamma_{^6 He} = \gamma_{^{18} Ne} = 120$.
The $^8$Li and $^8$B fluxes are tuned at the first oscillation peak, whereas
the $^6$He and $^{18}$Ne fluxes are tuned at the second oscillation peak.
A given ion is accumulated in the storage ring for five years.

{\bf Standard ``medium'' $\gamma$ scenario} $L = 650$ Km (CERN to Canfranc); $\gamma_{^6 He} = \gamma_{^{18} Ne} = 350$.
Both fluxes are tuned to be at the first oscillation peak.
A given ion is accumulated in the storage ring for ten years.

In all scenarios, a $\bar \nu_e$ flux of $2.9 \times 10^{18} \bar \nu_e$ per year or
a $\nu_e$ flux of $1.1 \times 10^{18} \nu_e$ per year is aimed at the distant detector, regardless of the decaying ion.
For a 5 year running time per each ion stored, the total luminosity is $1.45 \times 10^{19}$ $\bar \nu_e$
and $5.5 \times 10^{18}$ $\nu_e$ aimed at the far detector. We will consider a 1 Mton mass water \v Cerenkov detector (500 Kton fiducial mass),
with a MEMPHYS design~\cite{Campagne:2006yx}. At the considered neutrino energies, this detector is believed to show
a rather good neutrino energy reconstruction capability, allowing for a significant background rejection.
Migration and background matrices for this detector have been taken from Ref. 18.

\section{Solving degeneracies}

\begin{figure}[t!]
\vspace{-0.5cm}
\begin{center}
\begin{tabular}{cc}
\hspace{-0.55cm} \epsfxsize7.5cm\epsffile{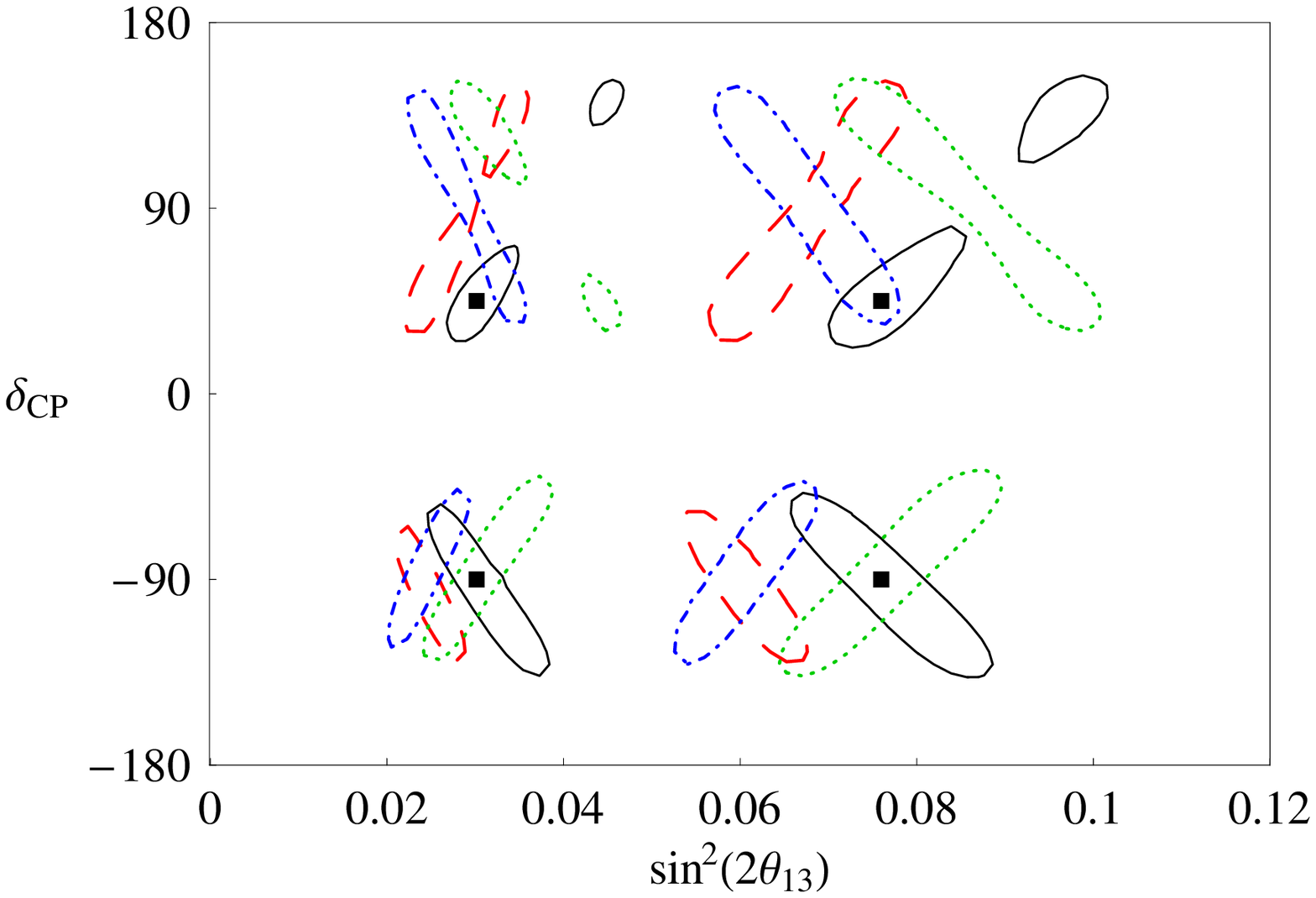} &
                 \epsfxsize7.5cm\epsffile{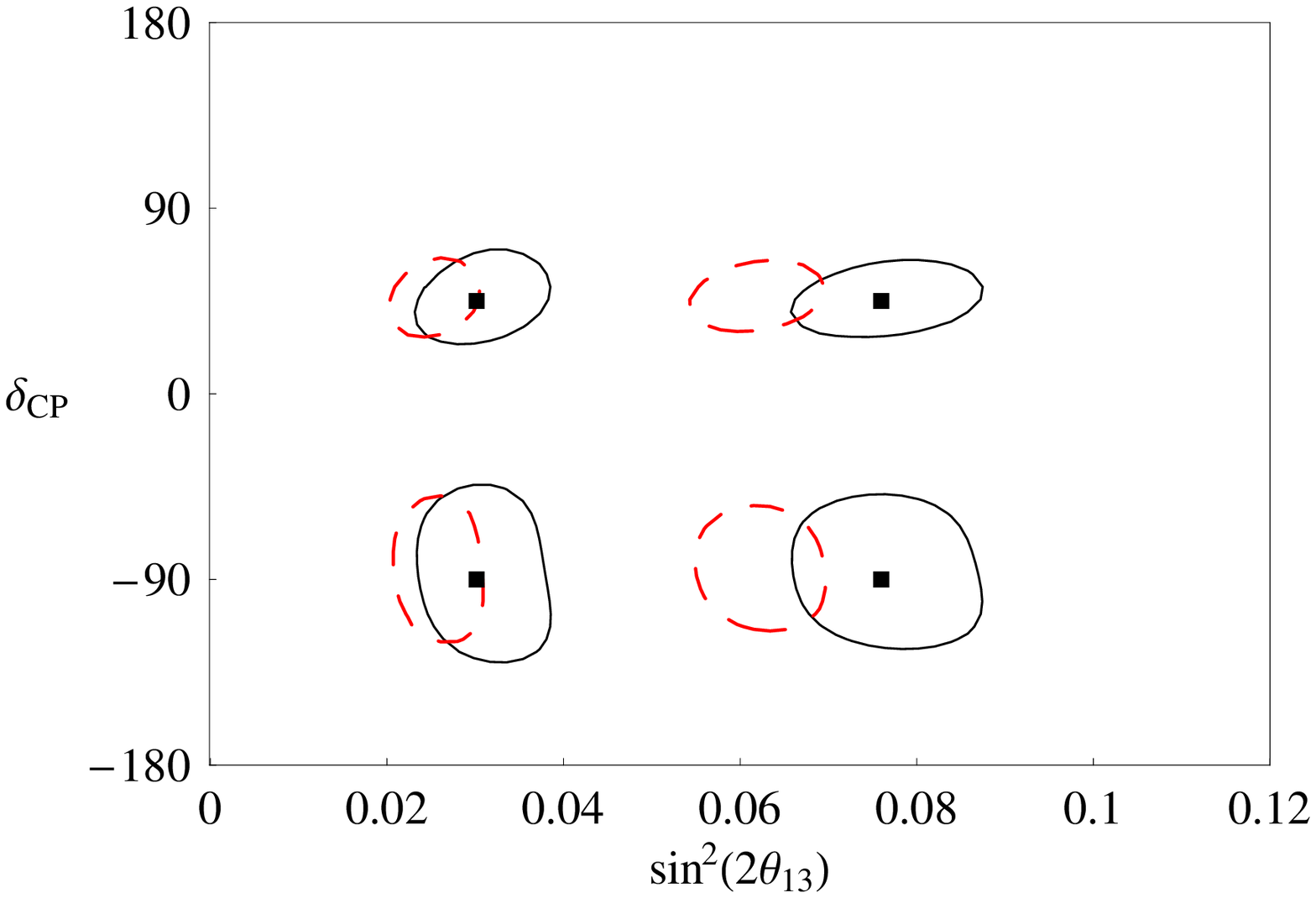}
\end{tabular}
\caption{\it
90\% C.L. contours for the standard ``low'' $\gamma$ setup (left)
and the alternating ions ``low'' $\gamma$ setup (right). The input pair is labelled by a thick black square
and corresponds to $\theta_{13} = 5^\circ, 8^\circ$ and $\delta = 45^\circ, -90^\circ$.
}
\label{fig:plots}
\end{center}
\end{figure}

In Fig.~\ref{fig:plots} we present our results for a fit to ``experimental data'' (generated as
in Ref. 2) for the {\bf standard ``low'' $\gamma$ setup} (left plot)
and the {\bf alternating ions ``low'' $\gamma$ setup} (right plot). The input pair is labelled by a thick black square.
Lines represent 90\% C.L. contours. The sign of $\Delta m_{23}^2$ has been chosen to be positive.
Since the sign of $\Delta m^2_{23}$ and the $\theta_{23}$-octant are unknown, fits to both sign[$\Delta m^2_{23}$] = $\pm$ 1 and
sign[$\tan (2 \theta_{23})$] = $\pm$ 1 have been performed (see Ref. 16 for a description of parametric degeneracies at the
``low'' $\gamma$ $\beta$-beam).

As it can be seen, for every considered input pair the {\it Alternating Ions Scheme} reduces the eightfold-degeneracy in the
($\theta_{13},\delta$) plane to a twofold-degeneracy. The so-called ``intrinsic'' degeneracy \cite{Burguet-Castell:2001ez}
is always solved (as it is usual when combining information from neutrino beams with different $L/E$).
Most importantly, the sign of the atmospheric mass difference is measured.
This is not possible at the standard ``low'' $\gamma$ setup since the baseline is too short
to take advantage of matter effects to discriminate between hierarchical and inverted spectra.

\begin{figure}[t!]
\vspace{-0.5cm}
\begin{center}
\begin{tabular}{cc}
\hspace{-0.55cm} \epsfxsize7.5cm\epsffile{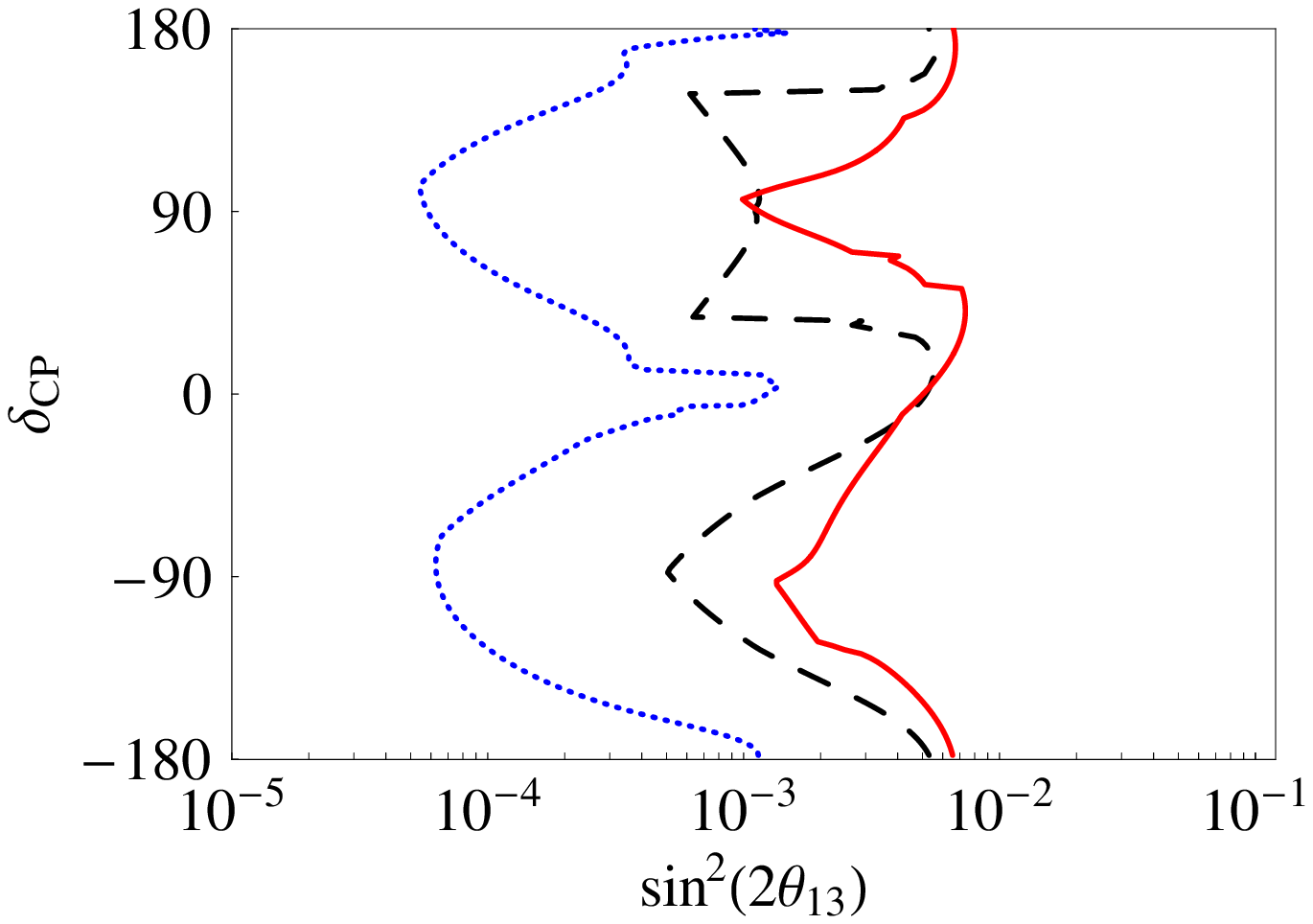} &
                 \epsfxsize7.5cm\epsffile{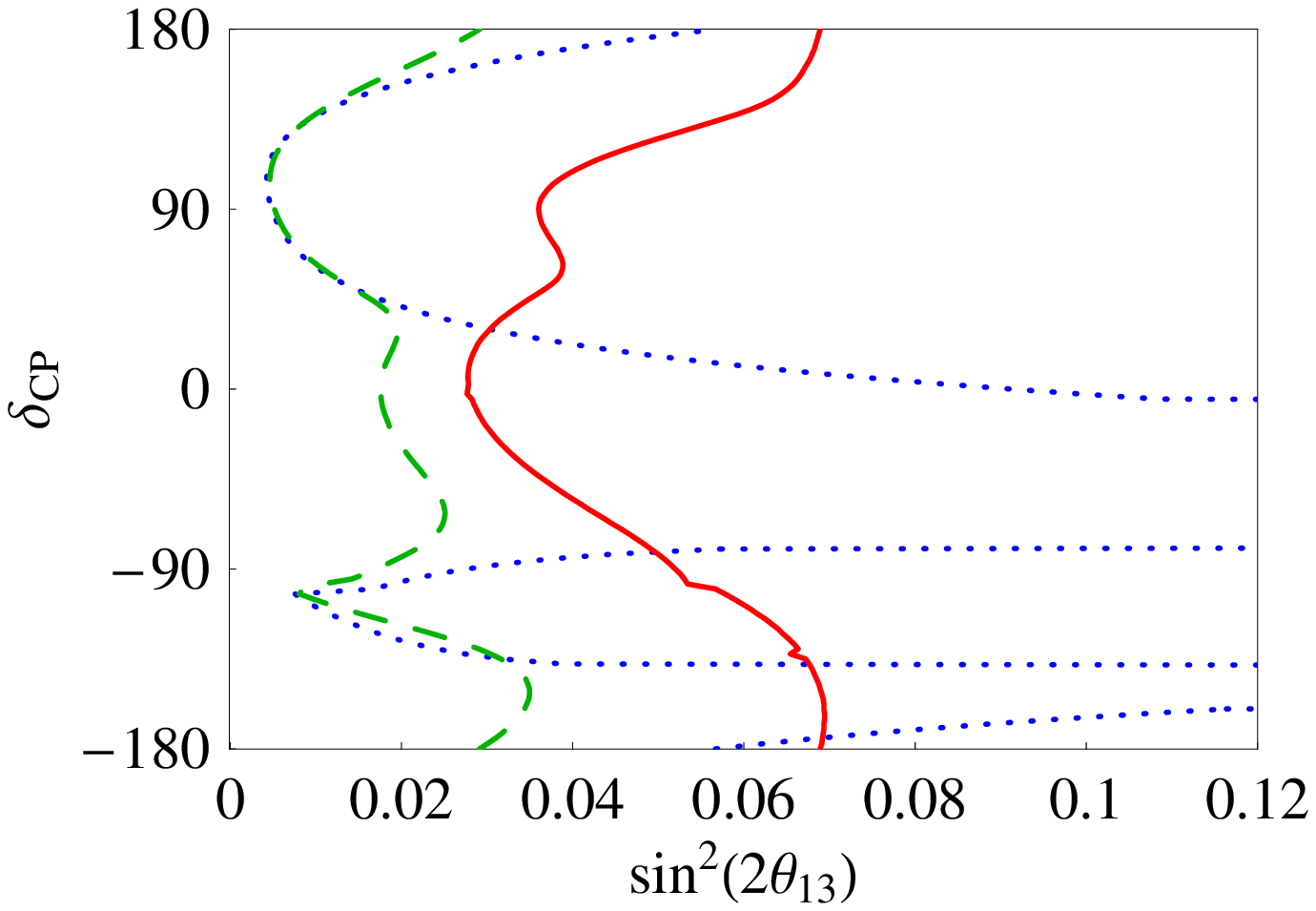}
\end{tabular}
\caption{\it
Comparison of the 3$\sigma$ $\theta_{13}$-sensitivity (left) and the sign($\Delta m^2_{23}$)-sensitivity
(right) for the alternating ions ``low'' $\gamma$ setup (solid) and
the standard ``medium'' $\gamma$ setup (dotted). The dashed line stands for the standard ``low'' $\gamma$ 
setup in the left panel and a combination of ``low'' and ``medium'' $\gamma$ in the right panel.
}
\label{fig:sens}
\end{center}
\end{figure}

In Fig.~\ref{fig:sens} we present our results for the $\theta_{13}$-sensitivity (left panel) and
sign($\Delta m^2_{23}$)-sensitivity (right panel) for the standard ``low'' $\gamma$ setup (dashed lines in the left panel),
the alternating ions ``low'' $\gamma$ setup (solid lines) and the standard ``medium'' $\gamma$ setup (dotted lines).
Notice that sensitivities are defined as in Refs. 16 and 22, taking into account
all of the parametric degeneracies. Lines represent 3$\sigma$ contours.
As it was expected, our {\it Alternating Ions Scheme} has the worst $\theta_{13}$-sensitivity as a consequence of the lower
statistics (Fig.~\ref{fig:sens}, left). The decrease in statistics due to the longer baseline is
compensated at the ``medium'' $\gamma$ setup by an increase in $\gamma$, providing a far better sensitivity than that of the
standard ``low'' $\gamma$ one, as noticed in Ref. 18.
On the other hand (Fig.~\ref{fig:sens}, right), the {\it Alternating Ions Scheme} can measure the sign of the atmospheric mass difference
for $\sin^2 (2 \theta_{13}) \geq 0.04$ ($\theta_{13} \geq 5^\circ$). This must be compared with the standard
``low'' $\gamma$ setup, with no sensitivity to sign($\Delta m^2_{23}$) due to its too short baseline.
Even compared to the standard ``medium'' $\gamma$ setup, with a much larger statistics, our scheme is particularly
effective for small $|\delta|$, a region of the parameter space in which other setups are not working
very well \footnote{This was the case for the Neutrino Factory, also, see Ref. 22.}.
This is a consequence of the combination of neutrino beams tuned at different oscillation peaks, something that
guarantees that ``sign clones'' are located in different regions of the ($\theta_{13},\delta$) parameter space
also in the case of a vanishing $\delta$. But this fact could also be exploited if higher $\gamma$ is available. As we have discussed
the ``medium'' $\gamma$ setup clearly outperforms the lower $\gamma$ ones in sensitivity to $\theta_{13}$ and to CP violation
for small $\theta_{13}$. Thus, if $\theta_{13}$ is not discovered at the next generation of experiments, such as T2K, the
``medium'' $\gamma$ setup would be the best choice to measure $\theta_{13}$ and $\delta_{CP}$. However, if nature has been unkind,
we could find that we are in a region of the parameter space on Fig.~\ref{fig:sens} in which this setup has no sensitivity to the sign.
An interesting possibility would then be to run the $\beta$-Beam at the lower $\gamma$, so that the oscillation probability is at its second peak for the same detector baseline, and combine it with the higher energy run at the first peak.
This is depicted by the dashed line in the right panel of Fig.~\ref{fig:sens}, where the regions in which the ``medium'' $\gamma$ setup could not
measure the sign are now also covered by the sensitivity curve. 

In summary, we think that the {\it Alternating Ions Scheme} is particularly well-suited to solve degeneracies in the physics case
where $\theta_{13}$ is relatively large (i.e. measurable at T2K-I). For smaller values of $\theta_{13}$ higher $\gamma$ factors are required but the degeneracies could also by softened in this case by combining runs with different $\gamma$ factors.


\section*{References}

\begin{thebibliography}{99}


\bibitem{Fogli:2005cq}
  G.~L.~Fogli {\it et al.},
  arXiv:hep-ph/0506083;
  G.~L.~Fogli {\it et al.},
  arXiv:hep-ph/0506307.


\bibitem{Cervera:2000kp}
A.~Cervera {\it et al.} ,
Nucl.\ Phys.\ B {\bf 579} (2000) 17 [arXiv:hep-ph/0002108].

\bibitem{Donini:2002rm}
A.~Donini {\it et al.},
Nucl.\ Phys.\ B {\bf 646} (2002) 321 [arXiv:hep-ph/0206034];
  D.~Autiero {\it et al.},
  Eur.\ Phys.\ J.\ C {\bf 33} (2004) 243
  [arXiv:hep-ph/0305185].

\bibitem{Burguet-Castell:2001ez}
J.~Burguet-Castell {\it et al.},
Nucl.\ Phys.\ B {\bf 608} (2001) 301 [arXiv:hep-ph/0103258].

\bibitem{Minakata:2001qm}
H.~Minakata and H.~Nunokawa,
JHEP {\bf 0110} (2001) 001 [arXiv:hep-ph/0108085].

\bibitem{Fogli:1996pv}
G.~L.~Fogli and E.~Lisi,
Phys.\ Rev.\ D {\bf 54} (1996) 3667 [arXiv:hep-ph/9604415].

\bibitem{Barger:2001yr}
V.~Barger {\it et al.},
Phys.\ Rev.\ D {\bf 65} (2002) 073023 [arXiv:hep-ph/0112119].

\bibitem{Itow:2001ee}
Y.~Itow {\it et al.},
arXiv:hep-ex/0106019.

\bibitem{Zucchelli:sa}
P.~Zucchelli,
Phys.\ Lett.\ B {\bf 532} (2002) 166;
  arXiv:hep-ex/0107006.

\bibitem{Geer:1997iz}
S.~Geer,
Phys.\ Rev.\ D {\bf 57} (1998) 6989 [Erratum-ibid.\ D {\bf 59} (1999) 039903] [arXiv:hep-ph/9712290];
A.~De Rujula {\it et al.},
Nucl.\ Phys.\ B {\bf 547} (1999) 21 [arXiv:hep-ph/9811390].

\bibitem{Donini:2006dx}
  A.~Donini and E.~Fernandez-Martinez,
  arXiv:hep-ph/0603261.

\bibitem{Rubbia:2006pi}
  C.~Rubbia {\it et al.},
  arXiv:hep-ph/0602032.

\bibitem{Burguet-Castell:2003vv}
  J.~Burguet-Castell {\it et al.},
  Nucl.\ Phys.\ B {\bf 695}, 217 (2004) [arXiv:hep-ph/0312068].

\bibitem{Bouchez:2003fy}
  J.~Bouchez {\it et al.},
  AIP Conf.\ Proc.\  {\bf 721} (2004) 37 
  [arXiv:hep-ex/0310059];
  M.~Mezzetto,
  J.\ Phys.\ G {\bf 29} (2003) 1771
  [arXiv:hep-ex/0302007];

\bibitem{Donini:2003vz}
A.~Donini {\it et al.},
JHEP {\bf 0406}, 011 (2004) [arXiv:hep-ph/0312072].

\bibitem{Donini:2004hu}
  A.~Donini {\it et al.},
  Nucl.\ Phys.\ B {\bf 710} (2005) 402
  [arXiv:hep-ph/0406132].

\bibitem{Donini:2004iv}
  A.~Donini {\it et al.},
  Phys.\ Lett.\ B {\bf 621} (2005) 276
  [arXiv:hep-ph/0411402].

\bibitem{Burguet-Castell:2005pa}
  J.~Burguet-Castell {\it et al.},
  Nucl.\ Phys.\ B {\bf 725}, 306 (2005)
  [arXiv:hep-ph/0503021].

\bibitem{Huber:2005jk}
  P.~Huber {\it et al.},
  Phys.\ Rev.\ D {\bf 73}, 053002 (2006)
  [arXiv:hep-ph/0506237];
  F.~Terranova {\it et al.},
  Eur.\ Phys.\ J.\ C {\bf 38}, 69 (2004) [arXiv:hep-ph/0405081].

\bibitem{Donini:2005qg}
  A.~Donini {\it et al.},
  arXiv:hep-ph/0604229.

\bibitem{Campagne:2006yx}
  J.~E.~Campagne {\it et al.},
  arXiv:hep-ph/0603172.

\bibitem{Donini:2005db}
  A.~Donini {\it et al.},
  arXiv:hep-ph/0512038;

\end{thebibliography}
\end{document}